\documentclass[aps,prl]{revtex4}
\usepackage{graphicx}
\usepackage{amssymb}
\usepackage{tikz-cd}
\def\ee{\end{equation}}
\def\bea{\begin{eqnarray}}

\begin{document}

\title{Collapse and Measures of Consciousness}

\author{Adrian \surname{Kent}}
\affiliation{Centre for Quantum Information and Foundations, DAMTP, Centre for
  Mathematical Sciences, University of Cambridge, Wilberforce Road,
  Cambridge, CB3 0WA, U.K.}
\affiliation{Perimeter Institute for Theoretical Physics, 31 Caroline Street North, Waterloo, ON N2L 2Y5, Canada.}
\email{A.P.A.Kent@damtp.cam.ac.uk} 

\date{September 2020; updated December 2020} 

\begin{abstract}
There has been an upsurge of interest lately in developing Wigner's
hypothesis that conscious observation causes collapse by exploring
dynamical collapse models in which some purportedly quantifiable aspect(s) of
consciousness resist superposition.    
Kremnizer-Ranchin, Chalmers-McQueen and
Okon-Sebasti\'an have explored the idea that collapse may 
be associated with a numerical measure of consciousness.
More recently, Chalmers-McQueen have argued that any
single measure is inadequate because it will allow superpositions
of distinct states of equal consciousness measure to persist. 
They suggest a satisfactory
model needs to associate collapse with a set of measures
quantifying aspects of consciousness, such as the ``Q-shapes''
defined by Tononi et al. in their ``integrated information
theory'' (IIT) of consciousness.   
I argue here that Chalmers-McQueen's argument against associating
a single measure with collapse requires a precise
symmetry between brain states associated with different experiences 
and thus does not apply
to the only case where we have strong intuitions,
namely human (or other terrestrial biological) observers.

In defence of Chalmers-McQueen's stance, it might be argued that idealized 
artificial information processing networks could
display such symmetries.   However, I argue that the most natural form
of any theory (such as
IIT) that postulates a map from network states to mind states is one that assigns identical
mind states to isomorphic network states (as IIT does).   
This suggests that, if such a map exists, no familiar components of mind states, such as viewing different colours, 
or experiencing pleasure or pain, are likely to be related by symmetries.  

\end{abstract}
\maketitle

\section{Introduction}

The hypothesis that wave function collapse is an objective process,
caused by conscious observation, is widely attributed to Wigner
\cite{wigner1961remarks}; a more detailed history, starting with
discussions by London-Bauer and
von Neumann, is given by Chalmers and McQueen \cite{cmdraft,cmtalks}.

It has recently been revived by
proposals  \cite{cmtalks,chalmersfqxi,cmdraft,kremnizer2015integrated,okon2016back,okon2018consciousness,osdraft} aimed at defining precise dynamical theories that combine ideas
for proposed objective quantifications of aspects of consciousness --
in particular Tononi et al.'s ``integrated information theory'' (IIT)
\cite{oizumi2014phenomenology} and related ideas -- 
with objective dynamical collapse models.   Dynamical collapse models
\cite{ghirardi1986unified,ghirardi1990markov} propose
parametrised stochastic differential equations that approximately reproduce 
pure unitary quantum evolution in one regime and approximately
reproduce the mathematical effect of the projection postulate, which characterises
the effects of measuring a quantum state, in another regime.
With appropriate choices of equations and parameters, they can
imply all the successful predictions of quantum theory in interference
and other experiments to date, while making testably different predictions in
possible (although generally technologically challenging) future
experiments.  They thus are testable candidate solutions for the quantum
measurement problem.   

The various proposals
\cite{cmtalks,chalmersfqxi,cmdraft,kremnizer2015integrated,okon2016back,okon2018consciousness,osdraft} 
have different features and are based on different assumptions.
It seems fair to say that all of them are projects in progress, and 
in each case it remains open to question whether a fully defined and
generally viable dynamical collapse model with all the desired features
will emerge, even in the non-relativistic limit.    
Chalmers and McQueen \cite{cmdraft}, whose work is our main focus
here, propose to model IIT (or some other such classical model of 
consciousness) within quantum theory by quasiclassical operators.
On the face of it this seems viable in regions, such as our own
environment, where physics can be described quasiclassically.
It seems hard, though, to 
extend it in a way that would combine naturally with a quantum 
theory of matter to produce a fundamental theory applicable in
all domains.       

Many other fundamental questions can be raised about the motivations for and viability of
this programme and of those it subsumes; many of these are carefully
reviewed in Ref. \cite{cmdraft} and in works cited therein.   
All versions of quantum theory with explicit collapse are somewhat 
ad hoc; relativistic collapse models are hard to define; Everettians
(see e.g. \cite{saunders2010many}) would argue that collapse models are unnecessary.     
Many (e.g. \cite{dennett1993consciousness}) argue that everything
about consciousness is completely explained by known science; on this
view, it seems no more plausible that it plays a role
in fundamental physics than that any other biological phenomenon does.        
In the other camp, many who see motivation for a theory of consciousness
criticise IIT as ad hoc, under-defined and having implausible implications
(see e.g. \cite{barrett2014integration,
aaronsonblog,
  cerullo2015problem, 
  barrett2019phi}).  
Still, as Chalmers-McQueen note, the idea that consciousness causes
collapse has some motivation, was taken seriously by some of
the pioneers of quantum theory, and, given that there is no consensus solution to the quantum measurement
problem, seems worth keeping on the table
for now and examining more carefully. \footnote{This might produce compelling
arguments against it; that too would be valuable.} 
As they also note, their proposals and arguments should apply to
a wide range of quantitative theories of consciousness; we may take 
IIT as a placeholder that illustrates some of the issues that would
also arise with possibly more satisfactory proposals.   

For the sake of discussion, let us accept this view.  This note focuses on one key question that
Chalmers and McQueen discuss: could a satisfactory conscious-collapse model
be defined in which the only quantity relevant to the collapse rate
is some proposed quantitative measure of consciousness, such as 
the $\Phi$ measure of IIT?   
The idea here is that ``$\Phi$ resists superposition and
superpositions of $\Phi$ trigger collapse''.   That is, a
superposition of
two or more quantum states that include a possibly conscious
subsystem, in which the superposition components contain
states of that subsystem that have different values of $\Phi$,
will undergo collapse at a rate that depends on the details
of the model and the relevant $\Phi$ values.   Moreover, 
this is the {\it only} cause of collapse, or at least the 
only cause generally relevant in situations where biological
organisms or classical or quantum computers observe -- and, in the
absence of collapse, would thus themselves enter and remain in -- quantum
superpositions. 

Chalmers and McQueen argue that \cite{cmtalks}
``$\ldots$ this view faces a fatal problem.'' 
Their concern is that it fails to suppress 
superpositions of qualitatively distinct but equal-$\Phi$ conscious
states.
For example, they consider \cite{cmpriv} an experiment in which a
conscious
subject observes a screen that can display blue or green in a dark
isolated room.
If the screen is put into a superposition of displaying both, then
the subject will be put into a superposition of both experiences.
They argue that there is no reason to assume that these experiences
differ in their $\Phi$ value.    If so, then there is no superposition
of states with distinct $\Phi$ values, and so a $\Phi$-collapse model
will leave the superposition uncollapsed.   
 

\section{Human observers}

It seems to me there {\it is} good reason to assume that the relevant
experiences differ in their $\Phi$ value, at least for terrestrial
animals -- the only subjects we have good
reasons to believe to be conscious.
Brains and central nervous systems are messy, noisy, imperfect
networks.   It seems very unlikely that, for any given subject,
the blue and green screens would excite exactly the same numbers
and rates of firing for retinal cells.   Even if they did, it
seems very unlikely that the subsequent chains of firings would
be along isomorphic neural network paths, at the same sets of
times.   And even if {\it this} were true in some impossibly
refined meditative state, in which no other brain processes
relevant to consciousness interacted in any way with the pure
perception of blue or green, it seems very unlikely that this
state could stably persist for any length of time.
The reason is that introspection and neural network models both
suggest we build up models of the world by association and memory.
When we see (say) blue, our dominant perception is of the colour, but it
is tinged with memories of landscapes, animals and art, with
past experiences, and with emotional associations with all of these. 
These may flicker in and out of consciousness, but are hard to 
suppress completely.  And even if we can fleetingly manage to focus
on the pure sensation of blueness, our unconscious neural processing
is still working on a network that encodes the learned associations,
and our $\Phi$ value depends (in general) on all these details.   
Our associations with and memories of green are different, and so
we should expect the associated $\Phi$ value to be.   
All of this, I suspect, is also true (to a greater or lesser, but
nonzero, extent)
for any pair of distinct colours that we can consciously distinguish.      

One possible response is that, even if not identical, the relevant
$\Phi$ values must be close.   However, this is too vague.
The $\Phi$ values may be close in the sense that the difference
between them is very small compared to the range of $\Phi$ values
that human brains can produce, but still distinct enough to imply
a swift (compared to human perception times) collapse in models
consistent with all available empirical data.  
To confirm or exclude this possibility, we would need a quantitative
estimate, presumably calculated from IIT applied to a (presently unavailable) 
neuron-level model emulating the observer's brain, together
with quantitative parameter bounds for $\Phi$-collapse models,
in order to calculate the possible range of collapse rates
for the green-blue superposed subject.    
 
It is helpful to compare the case of proposals 
for gravitationally induced collapse
\cite{penrose1996gravity,diosi1987universal}.   These suggest that collapses
of superpositions of matter states with distinct gravitational
fields take place very fast, even when the relevant fields
are almost identical by ordinary laboratory scale measures.   On Diosi's and Penrose's
estimates, a superposition of distinct mass distributions should
collapse long before the
relevant gravitational fields could be directly distinguished
by any laboratory measurement, for example. 
It seems quite natural to imagine that almost imperceptible
differences in levels of consciousness could similarly cause 
swift collapse in viable models.    

In fact, experimental evidence \cite{donadi2020} recently appears to 
have excluded the most natural proposal \cite{penrose1996gravity}
(so far) for a quantitative estimate of the rate of gravitationally
induced 
collapse.   
This only reinforces the point that we need parametrised 
models and empirical bounds to draw clear conclusions about
viable consciousness-collapse models, just as for any other
form of dynamical collapse model.
It should be noted that Diosi, the other authors of
Ref.\cite{donadi2020} and Penrose all continue to find the hypothesis of
gravitationally induced collapse attractive and natural, while
accepting that it may require a different formulation.  Whether or not it
will ultimately prove theoretically or empirically justifiable, the intuition
remains that collapse should be swiftly induced for superpositions
of states whose associated gravitational fields are very
similar by any directly observable measure.
If, as Chalmers-McQueen argue \cite{cmdraft}, consciousness-induced
collapse models deserve to be on the table at all, the analogous
intuition for these models seems similarly natural.   Of course, 
it may be refutable
(possibly even from existing data)
for IIT-based or other specific collapse models, but I 
see no reason to dismiss it without quantitative arguments. 

Another possible response is that, while the states of observing blue
screen and green screen may produce different $\Phi$ values,
we could tune them to produce two states with the same
$\Phi$ value.   The thought here is that $\Phi$ should
(at least to very good approximation) 
vary continuously with controllable parameters such as
screen size or intensity.    
So, if, say, the blue state has higher $\Phi$ value
than the green state, we could intensify or enlarge 
the green screen, until the $\Phi$ values are equal. 
Supposing this is correct, one problem is that we 
still would not know {\it which} amplified green
state has the same $\Phi$ value as the blue state,
and so we would not be able to {\it knowingly} create
a superposition state -- again, unless we had a 
neuron-level model emulating the observer's brain
and allowing us to calculate directly the relevant
$\Phi$ values.     We could try a range of amplifications,
in the hope that one of them is right -- but again,
without quantitative calculations and quantitative
models we cannot be sure any given set has significant probability of
producing a superposition of states with identical $\Phi$ values,
or of states with close enough $\Phi$ values to produce a 
long-lived superposition.
We could turn a dial continuously, amplifying the green
screen from much below the intensity of the blue screen
to much above.  At some instant, one might think this should produce equal
$\Phi$ values.   One problem with this is that $\Phi$ depends on
network transitions, and when a dynamic image's trajectory includes
a given image, the $\Phi$ value of the former may not necessarily be
close to that created by a static version of the latter.
Another is that the argument needs a superposition of states whose
$\Phi$ values remain equal for a significant length of time.    

In any case, we still have the problem of $\Phi$ instability arising
from instability of the content of consciousness, because of the vagaries of the human brain.    
Even if one of these methods produces a superposition of 
identical $\Phi$ states, for suitably tuned blue and green screen
observations, at a given point in time, the states will not 
persist as pure perception states and their $\Phi$ values 
should not be expected to remain identical for any significant
time.   

\section{Ideal observers}

What, though, about computers or other artificial observers?    
While we may not have very strong reasons to assume that they have
conscious states, IIT -- which can assign high values of $\Phi$ even
to quite simple computing devices -- suggests that they do.   
Perhaps this is true of any plausible theory that quantifies consciousness as
a function of information flow in networks.   We can certainly design
abstract networks that behave in precisely similar ways in response
to different inputs.   For example, we could make a network with
a detector array whose individual detectors generate a $0$ every
second if they receive blue light above a threshold intensity and
a $1$ if they receive green light above that intensity, and then
send these signals to two separate identical sub-networks that 
process them and characterise the blue or green shape detected.     
IIT suggests a suitably designed network of this type would be conscious 
for either input, with the same value of $\Phi$ in each case.
However, IIT postulates \cite{balduzzi2009qualia} identical conscious states associated with  
isomorphic networks. \footnote{I thank Kelvin McQueen for helpful
  discussions on this.} 
A superposition state of the network observing blue and green
screens would not collapse, but both components would
be associated with the same conscious state (whatever it is).  

Still, it is surely possible to find networks that produce precisely the
same value of $\Phi$ in two different non-isomorphic states,
to which IIT would assign distinct conscious states.  
A $\Phi$-collapse model would predict that
such a network {\it could} be put into 
a persistent superposition of conscious states.   
Is {\it this} a good reason to reject $\Phi$-collapse models?

To argue that it is, one has to assume that persistent superpositions
of distinct conscious states are generally unacceptable, even in 
cases very far from our own experience.    The thought here 
would presumably be that such states are just nonsensical
or uninterpretable, or at least that it is eminently reasonable to postulate some fundamental
principle that excludes them, perhaps following a loose analogy with
Penrose and Diosi's suggestion that nature does not allow superpositions of distinct
spacetimes.    

One might try to argue that if {\it any} persistent superpositions of distinct conscious 
states {\it were} straightforwardly intelligible, then purely unitary
quantum theory would be straightforwardly intelligble, and there would
be no motivation to consider any form of collapse.  
Arguments like this (e.g. \cite{bassi2010breaking,kent2018perception}) are used to justify 
lower bounds on collapse rates in dynamical collapse
models, but in that context they apply only to humans, relying on our 
introspective impression that we quickly see definite outcomes to 
a priori uncertain quantum experiments. 
To run a more general version of the argument one would need to show
that any sensible interpretation of persistent superpositions of
distinct conscious states for general (not necessarily human) observers would necessarily explain the appearance of single
worlds governed by Born rule probabilities for human observers in purely
unitary quantum theory.    

However, there are other possibilities that seem logically coherent.  For example,
one could imagine a psychophysical principle according to which
consciousness disappears in any persistent significant superposition of states that, individually,
would correspond to distinct conscious states.   (Here ``significant''
is a placeholder for some quantitative criterion.)   On such a rule, our ideal network
would ``be in conscious state $A$'' or ``be in conscious state $B$'' given the corresponding inputs, but 
have no experience when given a superposed input.   If humans cannot sustain persistent superpositions of
equal-$\Phi$ states (even in the absence of any collapse postulate)
because the $\Phi$ values of our complex and noisy brains
continually vary whatever their initial state, we would never have
encountered this effect. 
Even if we managed to sustain a superposition momentarily, it would
swiftly become a superposition of unequal-$\Phi$ states and collapse. 
We should expect this to leave no memory of the effect, for at least
two reasons.   First, the effect may be so fleeting as to be almost
imperceptible.   Second, we are assuming some form of extension of
IIT's model of consciousness, according to which brain states cause
conscious states but there is no independent causal effect in the
other direction.  Given this, any post-collapse memories are defined
by the post-collapse brain state, and so would be of definite 
conscious states corresponding to one or other component of the superposition.
 
It seems then that, to preclude superpositions of equal-$\Phi$ states
for artificial devices, we simply have to assume (a) that a device in a superposition
of such states must necessarily be conscious, (b) there is no
conceivable sensible account of what its conscious state could be 
that does not undercut the motivation for considering collapse models
at all.       
These aren't ridiculous assumptions -- a thought underlying (a) might
be that if individual states carry consciousness then any superposition
should (perhaps because of some loose analogy with charge or
mass, or because ``being conscious'' should behave like a binary quantum
observable) and a thought underlying (b) might be that talking about a superposition of conscious
states is just a category error -- but they don't seem completely
compelling.  

We should note too that there might turn out to be 
interesting theories of consciousness that (unlike IIT) assign 
consciousness only to biological systems.   For any such theories,
there are no ideal observers, and so no way to create equal-$\Phi$ states.
More generally, this is true for any theories whose measure for all physical systems
has similar features to IIT's $\Phi$ for biological systems, in that
it is always unstable and 
will have different time evolutions in different states where it is
initially equal. 

\section{Qualia and Symmetries}

It is worth considering more carefully the intuition that humans 
are (at least to very good approximation) equally conscious viewing different coloured screens
in an otherwise darkened room. 
There is a weak version of this intuition: in either case the human is
awake, and all waking conscious states should have very similar levels of $\Phi$. 
If one only holds this weak version, then the argument has just
as much force if we consider even radically qualitatively different
waking states -- viewing the blue screen, stroking a hamster,
listening to Bach, and so on.    
The complex definition of $\Phi$ in IIT suggests, though, that
radically different waking states should generally have different
values of $\Phi$.   If IIT is correct and our introspective intuitions
also are, the differences should be smaller
than the difference between any of them and a drowsy or dreamless
sleeping state.   It may also be hard to predict, from introspection,
how the $\Phi$ levels of the waking states are ordered.  
Still, a $\Phi$-collapse will predict that superpositions of 
these states collapse, perhaps (depending on the quantitative
details) very swiftly.   
If it were possible to produce a ``cleaned-up'' artificial human
emulator that can sustain stable $\Phi$ values and can reproduce 
pure versions (without the complicating associations) of the human experiences  
of viewing a blue screen, stroking a hamster and listening to 
Bach, we should still expect the $\Phi$ levels to be different 
and so still expect $\Phi$-collapse.     
The weak version of the intuition thus does not seem able to 
support Chalmers-McQueen's argument, even for idealized human
emulators. 

It seems to me that the argument feels initially plausible because
it relies on a stronger version of the intuition: that watching
a blue screen and a green screen are qualitatively identical,
up to interchanging the colours, and that images of the same area and
intensity in different colours should evoke the same $\Phi$.  
This in turn seems to rely on the intuition 
that there is a natural relation -- a colour flip map -- between the conscious states evoked.    However they are
represented mathematically, they have the same structure: a shape or
a collection of pixels of the same colour.    For example, if consciousness is
a collection of qualia, we can map the
blue-screen conscious state to the green-screen one simply by
replacing every blue quale with a green quale, and vice versa.
If consciousness has some other structure, we can find an analogous map with the same
effect.    
While the complexities of human brains may make the stronger intuition
not quite precisely correct for us, this intuition suggests hat the symmetries
should be exact rather than approximate for an ideal (artificial) observer that is able to have stable and
pure colour experiences. 

In any theory of consciousness that maps physical states to mental
states, this intuition seems to require an associated map acting on the physical brain states, 
which also defines an (approximate) symmetry of physical states. 
One way to see this is to ask what we would conclude
if we learned that observing blue and green screens used radically
different and unrelated brain pathways.   I think we would come to doubt
not only that the mind states are associated with similar levels of
consciousness, but also that they are as closely related as
introspection seems to suggest.

We can flesh all this out as postulating a commutative diagram:

\begin{equation}\label{commdiag}
  \begin{tikzcd}
{\rm Degree ~of~ consciousness} \arrow{r}{=} & {\rm Degree ~of~
  consciousness} \\
    {\rm Network ~states} \arrow{u}{\Phi}\arrow{r}{\rm effective~colour~flip}
    \arrow[swap]{d}{ Q} & {\rm Network ~states} \arrow{d}{Q} \arrow{u}{\Phi} \\
      {\rm Conscious ~states}  \arrow{r}{\rm colour~flip} & {\rm Conscious ~states} 
  \end{tikzcd}
\end{equation}

Here $Q$ is the map that characterizes the content and structure of
any consciousness associated to any given information processing
network: in IIT it is supposed to be representable mathematically
by the so-called qualia shape or Q-shape.   The ``effective colour
flip'' on network states maps a network state that evokes one
colour to a network state that evokes another; that is, it has
the effect of flipping the colour evoked in the conscious state. 

Why would one expect an effective colour flip map on network states
to give the equality in the $\Phi$ values in the top line? 
The third line {\it motivates} this, but to {\it justify} it
we need to consider the properties of the network model and
the hypothesized map $\Phi$.   
In the concrete example of IIT, a network $\times_i A_i = A_1 \times
\ldots \times A_n$ begins
at time $0$ in some initial state $a_1 (0) \times \ldots \times a_n (0)$, and at
each subsequent time step $t$ its state $a_1 (t) \times \ldots \times a_n (t)$
depends (in general probabilistically) on the previous state 
according to specified causal rules, which for fixed networks
are taken to be time-independent.   The degree of consciousness
$\Phi (t)$ at time $t$ is determined by the state at time $t$
and the rules determining the state at time $t+1$. 
The intuition that there is an effective colour flip map 
is mathematically natural (i.e. does not require fine-tuned
coincidences) only if there  is an associated permutation
symmetry $\rho$ such that the network $\times_i A_i$ in initial
state $\times_i a_i (0)$ is equivalent to the network
$\times_i A_{\rho(i)}$ in initial state $\times_i a_{\rho(i)}$, 
in the sense that the causal rules map isomorphically, so
that the probability of any future state $\times_i a_i (t) $ in
the former equals that of $\times_i a_{\rho(i)}
(t) $ in the latter.
This is also the natural physical justification: blue and 
green screens produce different inputs whose processing
paths through the network are isomorphic.  

But then the intuition fails: IIT does not assign different conscious states to
isomorphic network states \cite{balduzzi2009qualia}.   Moreover, this is not a peculiar feature of
IIT: it seems natural for any theory
mapping networks to conscious states.   
If the symmetry is precise, what could there be about $\times_i A_{\rho(i)}$ that could give
it the conscious state of green-screen perception where
$\times_i A_i$ gives blue?   The only difference is in the physical locations of the
relevant network components.   It seems very unnatural to suppose
that conscious qualia depend on these at all, and even more unnatural
to hypothesize (as one would have to) some fixed map from the
network's location in configuration
space to colour perception space.   Any choice of such a map would be
completely arbitrary: why should, say, a translation $1$mm to the left
swap blue with green?  And any permutation of colour qualia would
appear to define an equally valid and equally (im)plausible map.  

It might be argued that a precisely symmetric network
is never physically realisable.   However carefully one tries
to build a symmetric network, its transition probabilities
will always vary slightly from the designed specifications,
in a way that breaks the symmetry.    Indeed, but if so, 
the corresponding $\Phi$ values should also differ at least slightly,
and we return to the discussion of the previous section.  

It might also be noted that, even if a physical network is
symmetric in its high-level behaviour (as encoded in the
network transition probabilities), its components will never
be identical.  So, they can always be distinguished by features other than
their location.   Again (within classical models, and so ignoring
the possibility of and also the issues raised by indistinguishable
quantum states) this is true.   But if these features are irrelevant
to the conscious state then they do not matter; if they are
relevant, then the network model of consciousness considered was inadequate,
and a deeper model including these features is needed.  In this last
case,
again, the values of $\Phi'$ (the measure of consciousness in the 
new model) should differ.

\section{Symmetries and mind states}

There is an intriguing general point underlying this argument.   Discussions
of consciousness are often framed in a way that seems implicitly to appeal
to some form of symmetry among conscious states, or at least leaves
open the thoughts (i) that there may be such a symmetry and (ii) this
makes the arguments more plausible.  For example, James' discussion \cite{jamesautomata} of the 
difficulty in understanding the evolutionary emergence of epiphenomenal 
consciousness given the strong correlations between pleasure (pain)
and evolutionary (dis)advantage looks cleanest if one can simply
take pleasure and pain to be positive and negative values of a single
scalar quantity, whose sign one could imagine being reversed.  
Arguments involving hypothetical beings whose consciousnesses are
identical to ours except that they experience altered spectra (say
with blue/green exchanged) similarly seem cleanest if 
colour sensations are not only considered as elementary components
of mind states, independent of any other qualia, but also thought
of as completely interchangeable.  

Introspection gets us only so far on this.   Certainly colour sensations 
{\it feel} as though they belong to a common class, different
from auditory sensations, or verbalised thoughts, or emotions 
(though they may evoke at least the last two).  But is it possible
to experience pure colour and nothing else?   For the reasons given earlier,
I'm not so sure.   And do elementary pure colour sensations even exist 
(whether or not they can be experienced apart from other qualia)?  
It feels hard to me to identify distinct components of colour sensations, even
when experiencing colours that are mixtures of primary colours.
But that doesn't seem to be a strong argument that colour sensations
-- even primary colour sensations -- aren't fundamentally composite.   
Is exchanging blue and green qualia more analogous 
to exchanging up and down quarks (an approximate symmetry)
or positive and negative charges (a better approximate symmetry)
or applying the PCT operator (an exact symmetry) in the material world? 
Or is it more like exchanging $CO_2$ and $N_2O$ -- an operation
which might look superficially like a symmetry, if one does not 
know the underlying structure of the world, but is not
one in any meaningful sense? 
Introspection doesn't seem to give clear answers.    

Are different colour perceptions evoked by near-isomorphic brain
states?   I don't think we know this either.   Even if they are, the
deviations of the states from isomorphism, although small, may be crucial rather
than incidental to the experiences evoked.   
So our knowledge about our own physical brain states also does not seem to offer strong
support for either a physical symmetry (between brain states
associated with different colour qualia) or a mental one (between different
colour qualia). 

The case for a pain/pleasure symmetry seems weaker still.
A pure colour sensation 
at least {\it seems} (if perhaps mistakenly) imaginable. 
It seems much harder to imagine a pure pain or 
pleasure sensation, untethered to any past or anticipated event, to
any more complex emotion, or to any region of the body.   I am also not sure that it makes sense to describe a
pain as a negative pleasure; pains and pleasures (and even different
types of pain and pleasure) 
feel qualitatively different, not quantitatively.   
 It's true that rational 
human behaviour is often modelled as an attempt to maximize a scalar
quantity, utility, that can take positive and negative
values.   But theoretical justifications for this (e.g. \cite{savage1951theory})
come from plausible axioms about what constitutes rational behaviour.
They do not require the hypothesis that our experiences can be qualitatively 
{\it characterised} by a utility measure.  

Finally, as noted above, the hypothesis that there are 
isomorphic states of an idealised brain that evoke distinct
pure colour qualia related by a symmetry is quite problematic.   If we postulate the 
symmetry, it should apply both to brain states (those 
associated with pure colour qualia) and mind states (those
experiencing pure colour qualia).
But we then need to break the symmetry again in order to
get a map from specific brain states (associated with some
specific colour qualia) to specific mind states (the experience
of those specific qualia).    This is not inconceivable in principle --
spontaneous symmetry breaking is a familiar phenomenon elsewhere
in physics.    But spontaneous symmetry breaking needs an
explanation -- for example, dynamics induced
by a potential with a degenerate ground state -- and we have
no substantive model here that could offer one.  It would seem rather
perversely baroque to postulate symmetries in both the material and
mental worlds, for which we have no good 
evidence, and then further suggest that some unknown symmetry breaking
mechanism in some unknown model explains how their actions come to be  
related by maps with the properties given in (\ref{commdiag}),
with the symmetry realised in both worlds but broken by $Q$. 
Again, similar comments apply to models involving isomorphic 
pleasure/pain states.  

In both cases, then, I suggest that the hypothetical symmetries cannot
naturally be accommodated within physical theories
of consciousness that aim to map brain states to mind states.
IIT usefully illustrates this, but the argument is independent of 
the details of IIT.   
If so, arguments that both assume some such theory is correct and appear to be implicitly
drawing some strength from a hypothetical symmetry need to be carefully examined, and
if possible reframed in a way that is explicitly independent of
symmetry hypotheses.\footnote{James argues against an epiphenomenal
theory of consciousness, which in my view describes the most natural
interpretation of IIT when taken as a fundamental theory.   He does not discuss a
brain-mind map in detail, nor suggest how a scientific theory of
the evolution of consciousness could overcome his argument.  
The relevance of our arguments to his discussion is thus contingent on
the form of the unspecified theory.  In any case, my hunch is that his argument
could be adequately reframed: a much more careful discussion is needed, but here
is a sketch.   I find it hard to 
accept that it is simply tautologous to say that activities good
or bad for our genetic survival are respectively pleasurable or
painful.   If we agree that the statements are more than
tautologies then we agree, I think, that in the present state
of our understanding we could imagine the world being
otherwise.   That is, the correlations with evolutionary (dis)advantage
need some explanation, which we presently do not have.   There
is an obvious naive explanation -- we have evolved minds and bodies
interacting so that our physical behaviour is generally
pleasure-seeking and 
pain-avoidant -- but it is incompatible
with epiphenomenalism.  This was James' essential point, and we do not
need to imagine a simple pleasure/pain symmetry defining a swap 
map to run this framing of the argument.}

\section{Earlier relevant work}

After circulating the first draft of this paper, my attention was drawn
to earlier work on symmetries and consciousness, including intriguing discussions by Hurley
\cite{hurley1998consciousness}, Lee \cite{lee2006experience}
and Chalmers\cite{chalmers2019three}.
Each of these examines the case of left-right symmetry, 
which perhaps illustrates mostly clearly the points at issue
here, and Lee's discussion \cite{lee2006experience} is particularly
relevant.   

This topic deserves an independent discussion, which I hope to give
elsewhere.
Here I just summarize the implications of the arguments above.
Suppose consciousness is described by a theory like IIT, which 
defines a mathematical map from a space of classical descriptions of brain/network states
to the space of associated mind/conscious states.   
Consider a brain/network with perfect left-right symmetry.
If the map associated distinct, symmetrically related brain/network
states to distinct mind/conscious states, it would effectively give
an intrinsic label on three dimensional coordinate systems,
distinguishing left-handed sets of coordinates from right-handed. 
While that is logically possible, it would be surprising, since
it would break a symmetry of classical physics.   (Even in
a model based on quantum states, it would be surprising, since
we know no reason why the violations of parity in quantum field theory  
should be relevant to consciousness.)    

The more natural possibility, we have argued, is that symmetrically
related brain/network states are associated to identical conscious
states.  A hypothetical creature with a perfectly left-right symmetric
neural network would not distinguish ``leftish'' and ``rightish'' 
sensations, although it might respond appropriately (and differently,
though symmetrically) to stimuli on the left and right. 
Approximately left-right symmetric creatures may distinguish 
between perception or proprioception qualia associated with events on their left and
right.   These may perhaps feel to them (as they do for us)
qualitatively similar and it may seem intuitively plausible to them
(as it perhaps does to us) that they are in some sense related.   
But, on the view we are arguing for, this intuition cannot be
made precise.  The states 
are not approximations to distinct but perfectly
symmetrically related qualia states.   Structural asymmetry is
crucial to the sense of left-right distinctness.  Eliminating 
the asymmetry (for example, by some continuous deformation of 
the neural networks, if that were possible) would also eliminate
the sense of distinctness.

\section{Summary}

We have argued that collapse models based on a single measure of consciousness seem
consistent with experience.   Thought experiments that attempt to
place humans in superpositions of distinct but equally conscious
states do not refute such models, because it seems unlikely that 
humans can sustain precisely the same level of consciousness in
any state for long, or that there are distinct human conscious states that remain
precisely equally conscious for long.   Thought experiments
involving artificial networks, which might -- according to some theories
of consciousness -- sustain distinct and
equally conscious states for long periods, do not refute the models
either, since we do not know what, if anything, these networks 
would experience in superposition.

Another argument that single-measure collapse models are plausibly consistent
with our experience has been suggested by Okon and
Sebastian \cite{okon2018consciousness,osdraft}, who discuss
the effect of decoherence (of the measurement device) on the
environment and hence on an observer's consciousness. 
In response, Chalmers and McQueen \cite{cmdraft} argue that
any effects of decoherence are screened off from the
observer's consciousness in the blue- and green-screen experiment:
they see only the screen.    One could press this further by 
arranging quantum experiments in which blue or green light
pulses are sent towards the observer's retina, without
any classical amplification.    
The discussion can be made clearer still by using the type
of ideal artificial observers considered above. 
Within an IIT model, these can easily be designed so that only the experimental
outcome affects their information processing network, rendering any
decoherence effects irrelevant.  
For these reasons, we believe our independent arguments are needed.
    
We have also argued that there are strong reasons to doubt
the intuition that we can, from introspection, identify experiences
that are likely to have equal measures on consciousness, according to some
sensible theory mapping physical states to mental states.   It relies implicitly on a notion of symmetry between
distinct mental states, which is hard to make natural in such a
theory.   On the one hand, a symmetry between mental states ought
to be associated with a symmetry between the associated physical
states.   On the other, physical states related by simple symmetries
seem naturally associated with identical mental states. 

We are not suggesting here that there is no motivation to consider
collapse models based on a detailed (multi-parameter) mathematical
description of the contents of consciousness.  
There are other reasons \cite{cmtalks,cmdraft}
why one might prefer such models: for example, that this may give
consciousness a causal role in nature that aligns with some intuitions. 
These are interesting lines of thought, but beyond our scope here. 

\section{Acknowledgements}
This work was supported by an FQXi grant and by Perimeter Institute
for Theoretical Physics. Research at Perimeter Institute is supported
by the Government of Canada through Industry Canada and by the
Province of Ontario through the Ministry of Research and Innovation.
I thank David Chalmers and Kelvin McQueen for circulating a draft of
their article \cite{cmdraft} and for very helpful
and enjoyable discussions, and Johannes Kleiner for helpful comments
on a draft.   

\section*{References}

\bibliographystyle{unsrtnat}
\bibliography{qualia2}{}
\end{document}